\documentclass[aps,
prB,
,twocolumn,
longbibliography,
amsmath,
amssymb]{revtex4-2}
 
\usepackage{graphicx}
\usepackage[normalem]{ulem}
\usepackage{soul}
\usepackage{dcolumn}
\usepackage{bm}
\usepackage{multirow}
\usepackage{esint}
\usepackage[utf8]{inputenc}
\usepackage{relsize}
\usepackage{color}
\usepackage{siunitx}
\usepackage[thinlines]{easytable}

\usepackage{siunitx}
\usepackage{relsize}
\usepackage{blindtext}
\usepackage[thinlines]{easytable}
\usepackage{bm}
\usepackage{lipsum}

\newcommand{\SHE}{\sigma_{\text{\tiny SH}}}
\newcommand{\OHE}{\sigma_{\text{\tiny OH}}}
\newcommand{\SNE}{\sigma_{\text{\tiny SN}}}
\newcommand{\ONE}{\sigma_{\text{\tiny ON}}}

\newcommand{\SHC}{\sigma_{\text{\tiny SH}}}
\newcommand{\OHC}{\sigma_{\text{\tiny OH}}}
\newcommand{\SNC}{\sigma_{\text{\tiny SN}}}
\newcommand{\ONC}{\sigma_{\text{\tiny ON}}}

\begin{document}

\preprint{APS/123-QED}

\title{First-principles theory of intrinsic spin and orbital Hall and Nernst effects in metallic monoatomic crystals}

\author{Leandro Salemi}
 \email{leandro.salemi@physics.uu.se}
\author{Peter M. Oppeneer}%
\affiliation{%
 Department of Physics and Astronomy, Uppsala University, P.O. Box 516, SE-75120 Uppsala, Sweden
 }%
 \hspace{2cm}

\date{\today}

\begin{abstract}
The generation of spin and orbital currents is of crucial importance in the field of spin-orbitronics. In this work, using relativistic density functional theory and the Kubo linear-response formalism, we systematically investigate the spin Hall and orbital Hall effects for 40 monoatomic metals. The spin Hall conductivity (SHC) and orbital Hall conductivity (OHC) are computed as a function of the electrochemical potential and the influence of the spin-orbit interaction strength is also investigated. Our calculations predict a rather small OHC in $sp$ metals, but a much larger OHC in $d$-band metals, with maximum values [$\sim 8000\,(\hbar/e)\Omega^{-1}{\rm cm}^{-1}$] near the middle of the $d$ series. 
Using the Mott formula, we evaluate the thermal counterparts of the spin and orbital Hall effects, the spin Nernst effect (SNE) and the  orbital Nernst effect (ONE). We find that the as-yet unobserved ONE is significantly larger ($\sim 10 \times$) than the SNE and has maximum values for group 10 elements (Ni, Pd, and Pt). Our work provides a broad overview of electrically- and thermally-induced spin and orbital transport in monoatomic metals.
\end{abstract}

\pacs{Valid PACS appear here}
\maketitle

\section{Introduction}
The ability to control, generate and detect spin currents is a central issue to design efficient spintronics devices. 
Spin-polarized currents have already become exploited in spin-transfer torque magnetic random access memories (STT-MRAM) (see, e.g., \cite{Bhatti2017}). In the past ten years the spin-orbit torque (SOT) has come in the research focus, when it was discovered that the SOT leads to very efficient magnetization switching in magnetic elements \cite{MihaiMiron2011,Liu2012}. This drew the attention to the generation and utilization of pure spin currents that are understood to be behind the efficient  switching caused by the SOT \cite{Manchon2019}. 

One of the key phenomenon in the efficient generation of pure spin currents is the spin Hall effect (SHE), where a transverse spin current arises from a longitudinal charge current \cite{Dyakonov1971,Sinova2015}. Initially predicted based on the assumption of extrinsic electron scattering \cite{Hirsch1999}, impurities-independent intrinsic SHEs in various systems such as p-type semiconductors \cite{Murakami2003} or a two-dimensional electron gas  with a Rashba-type spin-orbit coupling (SOC) \cite{Sinova2004} were later theoretically predicted. The intrinsic contribution originates from the Berry curvature associated to the band structure of the material \cite{Murakami2003,Sinova2004,Tanaka2008, Xiao2010} while extrinsic mechanisms, such as skew-scattering and side jumps, originate from spin-dependent scattering on defects \cite{Smit1958,Berger1970}. It also became clear, via indirect measurements of the inverse SHE (ISHE), that metals exhibit orders of magnitude larger spin Hall conductivities (SHC) than semiconductors \cite{Saitoh2006,Kimura2007,Valenzuela2006}.

Extended studies on metallic alloys \cite{Hong2018,Zou2016,Wu2016,Wen2017,Laczkowski2014,Laczkowski2017,Qu2018,Zhu2018,Ramaswamy2017,Nguyen2016,Chen2017,Niimi2011, Niimi2014, Niimi2012} have showed that, while some systems mainly show extrinsic mechanisms \cite{Niimi2012, Ramaswamy2017, Laczkowski2014} others are mainly dominated by intrinsic contributions\cite{Nguyen2016, Zhu2018, Laczkowski2014}. The crossover between extrinsic and intrinsic contribution have been recently investigated in Au/Cu alloys, where varying the concentration of Cu can lead to either intrinsic-dominated or extrinsic-dominated SHE \cite{Musha2019}.

Further support for the existence of a large intrinsic SHE in metals came from \textit{ab initio} calculations of the SHC in Pt  that predicted a huge SHC $\sim 2000\, (\frac{\hbar}{e}) (\Omega\,\textrm{cm})^{-1}$ \cite{Guo2008}. This quickly raised the question of the physical origin of this huge SHE in metallic systems. Kontani \textit{et al.} \cite{Kontani2007} studied the particular case of Pt and using a tight-binding Hamiltonian and found that there exists in fact a huge orbital Hall effect (OHE) which arises from a phase factor analogous to the Aharonov-Bohm phase factor, without requiring any SOC. This phase factor is induced by an effective magnetic flux due to the angular dependence of the $d$-orbitals. Other theoretical works on the OHE were conducted on various systems such as 4$d$ and 5$d$ transition metals \cite{Tanaka2008, Kontani2009}, Sr$_2$MO$_4$ (M = Ru, Rh, Mo) \cite{Kontani2008} and heavy-fermion systems \cite{Tanaka2010}. 

Recently, a renewed interest has emerged for the OHE as it might offer an intriguing way to generate orbital currents and utilize these to perform magnetization switching \cite{Go-Lee2020}.  Recent works re-investigated the OHE in $d$-transition metals systems as well as $sp$ metals such as Li or Al \cite{Jo2018,Go2018}. These reached similar conclusions: the intrinsic OHE is a generic quantity that is bigger than the SHE and arises without the interplay of SOC while the intrinsic SHE arises as a result of the OHE and SOC. 

Those works offer new perspective on orbital-related phenomena, which often tend to be neglected due to the well known orbital-quenching in periodic solids. Other recent investigations  in closely related fields support the notion of orbital-driven physics in more exotic systems such as antiferromagnets \cite{Salemi2019} or in chiral structures \cite{Yoda2018}. Efforts are actually being devoted to unify spin and orbital dynamics for a more accurate modeling of SOTs \cite{Go2020c}.

The analogy between transverse charge and spin transport in the normal Hall effect and SHE has been extended to thermally-driven transport in the last years. Thermally-driven spin transport phenomena such as the spin-dependent Seebeck effect \cite{Uchida2008,Jaworski2010,Uchida2010} and spin Peltier effect \cite{Flipse2012,Flipse2014} have been observed. 
Transverse thermally-induced charge transport, given by the von Ettingshausen-Nernst effect \cite{Nernst1886}, was predicted to have a spin analog, the spin Nernst effect (SNE) \cite{Cheng2008}. The SNE was initially investigated theoretically \cite{Wimmer2013,Tauber2012,Liu2010b,Dyrdal2016}, and, recently, it was observed in Pt and W thin films \cite{Meyer2017,Sheng2017,Bose2018}.  

The SNE can be seen as the thermal counterpart of the SHE. Pushing the analogy a bit further, we propose the concept of the orbital Nernst effect (ONE), where a longitudinal temperature gradient induces transverse orbital angular momentum thermal transport. There is however not much known about the behavior of the SNE in metals and even less about the ONE.

To analyze trends in electrically- and thermally-driven transverse spin and orbital transport, we compute the intrinsic contribution of SHE, OHE, SNE, and ONE in 40 monoatomic crystals.

Our calculations are performed within the framework of relativistic density functional theory (DFT) and Kubo linear-response theory. The main focus in set on $3d$, $4d$, and $5d$ transitions metals but metals from the first and second columns (e.g., Li, Na) as well as $sp$ metals (e.g., Al, In, and Pb) are also considered. Compared to previous works which used tight-binding Hamiltonians \cite{Kontani2007,Tanaka2008, Kontani2009, Kontani2008,Tanaka2010,Jo2018,Go2018}, the use of all-electron, full-potential relativistic DFT allows in general for a more precise description of the electronic structure.

This paper is organized as follow. First, we introduce the concepts of spin and orbital currents in Sec.\ \ref{Theory}, looking at electrically and thermally driven transport coefficients. The Mott formula is used to link those two. The quantum description of the systems is discussed and the linear response formula is presented. Second, using our previously defined transport coefficients as well as the linear response framework, we compute in Sec.\ \ref{Results} the SHC and orbital Hall conductivity (OHC) as a function of the electrochemical potential for the 40 monoatomic crystals considered, with a specific focus on the $3d$, $4d$, and $5d$ series. The influence of SOC strength is investigated. Finally, we discuss the SNC and ONC throughout the elements considered. As we show later, the ONC is found to be the dominating quantity, being about 10 times bigger than the SNC.

\section{Theory}
\label{Theory}
\subsection{Spin and orbital transport}
The spin current density $\bm{J}^{S^k}$ is a $3$-dimensional vector describing the flow of spin angular momentum polarized along the $k$-direction ($k=x,y,z$) and can be related to the the external electric field $\bm{E}$ as
\begin{equation}
\label{eq:electrical_spin_transport}
J^{S^k}_i = \sigma^{S^k}_{ij} E_j, \hspace{0.5cm} (i, j = x,y,z)
\end{equation}
where we have assumed the Einstein summation notation. The quantity $\sigma^{S_k}_{ij}$ is the $ij^{\rm th}$ component of the {$2^{\text{nd}}$-rank} spin conductivity tensor $\boldsymbol{\sigma}^{S^k}$. Analogously, we can define the orbital current density $\bm{J}^{L^k}$ as well as the orbital conductivity tensor $\boldsymbol{\sigma}^{L^k}$ with
\begin{equation}
\label{eq:electrical_orbital_transport}
J^{L^k}_i = \sigma^{L^k}_{ij} E_j.
\end{equation}

In the presence of a thermal gradient, thermally induced spin and orbital flow can also occur. Extending Eqs.~(\ref{eq:electrical_spin_transport}) and (\ref{eq:electrical_orbital_transport}), we can write
\begin{align}
\label{eq:elec_thermal_spin_transport}
J^{S^k}_i &= \sigma^{S^k}_{ij} E_j - \Lambda^{S^k}_{ij}  \frac{\partial}{{\partial}r_j} T ,\\
\label{eq:elec_thermal_orbital_transport}
J^{L^k}_i &= \sigma^{L^k}_{ij} E_j - \Lambda^{L^k}_{ij}  \frac{\partial}{\partial r_j} T ,
\end{align}
where $\Lambda^{S^k}_{ij}$ ($\Lambda^{L^k}_{ij}$) is the $ij^{\rm th}$ component of the $2^{\text{nd}}$-rank spin (orbital) magneto-thermal conductivity tensor $\boldsymbol{\Lambda}^{S^k}$ ($\boldsymbol{\Lambda}^{L^k}$), and $T$ the temperature. Note that the electrical term of Eqs.~(\ref{eq:elec_thermal_spin_transport}) and (\ref{eq:elec_thermal_orbital_transport}) can be expressed in a similar way as the thermal term using the Cartesian spatial derivative of $E_j = -\frac{\partial}{\partial r_j} V$, where $V$ is the electric potential.

Thermally induced transport can be related to electrically induced transport 
through
\begin{equation}
\label{eq:Mott_Formula}
\boldsymbol{\Lambda}^{S^k(L^k)} = \frac{\pi^2 k_B^2 T}{-3e} \Big( \frac{d}{dE} \boldsymbol{\sigma}^{S^k(L^k)} \Big)_{\! E=E_F},
\end{equation}
where $k_B$ is the Boltzmann constant and $e$ the elementary charge. Equation (\ref{eq:Mott_Formula}) is the famous Mott formula \cite{Mott1969} applied to metallic systems. The derivative on the right-hand side is taken with respect to the electronic potential $E$ and evaluated at the Fermi energy $E_F$. Explicit calculations have proven that the Mott formula gives exact results for small temperature excursions \cite{Jonson1980}. Note that our focus here is on ``pure" thermal transverse spin and orbital transport. An additional effect can occur when a  longitudinal thermal electric current, described by the Seebeck coefficient $S$, is converted to a transverse spin (or orbital) current through the SHE (or OHE), which will give a contribution $\sim \sigma_{ij}^{S^k(L^k)}S\partial T / \partial r_j$ with $i$, $j$, and $k$ all different. This effect is not considered here, but it can be evaluated from the calculated transverse spin (orbital) conductivities and literature values for the Seebeck coefficient.

The expressions (\ref{eq:elec_thermal_spin_transport}) and (\ref{eq:elec_thermal_orbital_transport}) for the spin and orbital current density contain both longitudinal and transverse transport quantities. Our focus is here on the \textit{transverse} conductivities. 
These are, for nonmagnetic metals, given by the spin Hall  and orbital Hall conductivities, $\sigma_{ij}^{S_k}$ and $\sigma_{ij}^{L_k}$, which are nonzero and identical for all indices such that the Levi-Civita tensor $\epsilon_{ijk} \neq 0$. 

We can thus express the tensorial SHE and OHE by a single spin Hall conductivity (SHC) $\SHC$ and an orbital Hall conductivity (OHC) $\OHC$, respectively, which can be defined as $\SHC \equiv \sigma^{S^z}_{xy}$, and $\OHC \equiv \sigma^{L^z}_{xy}$.
In a similar way, the SNE and ONE can be related to their respective spin Nernst thermal conductivity (SNC) $\SNC$ and orbital Nernst conductivity (ONC) $\ONC$, both obtained using Eq.~(\ref{eq:Mott_Formula}), i.e.,
\begin{align}
\label{eq:SNC_Definition}
\SNC &\equiv  \frac{\pi^2 k_B^2 T}{-3e} \Big( \frac{d}{dE} \SHC \Big)_{E=E_F},\\
\label{eq:ONC_Definition}
\ONC &\equiv \frac{\pi^2 k_B^2 T}{-3e} \Big( \frac{d}{dE} \OHC \Big)_{E=E_F}.
\end{align}
The \textit{ab initio} calculation of the SHC and OHE is detailed in the following subsection. 

\subsection{First-principles linear response}

To evaluate the response quantities we consider the influence of an external electric field $\bm{E}$ which leads to an additional term $\hat{V}$ to the unperturbed Hamiltonian $\hat{H}_0$ that can be written as $\hat{V} = -e \hat{\bm{r}} \cdot \bm{E}$, where $e$ is the electron charge and $\hat{\bm{r}}$ the position operator. For the unperturbed Hamiltonian we adopt the relativistic Kohn-Sham Hamiltonian $\hat{H}_0$. Using the DFT package WIEN2k \cite{Blaha2018}, we solve the eigenvalue equation $\hat{H}_0 |n\bm{k} \rangle = \epsilon_{n\bm{k}} |n\bm{k} \rangle$, where $|n\bm{k} \rangle$ is the single-electron state at band-index $n$ and reciprocal wavevector $\bm{k}$ with eigenenergy  $\epsilon_{n\bm{k}}$. 
To compute the spin and orbital conductivity tensors, we first define their respective quantum mechanical operators $\hat{J}^{\hat{S}^k}_{i}$ and $\hat{J}^{\hat{L}^k}_{i}$ as
\begin{align}
\hat{J}^{\hat{S}^k}_{i} &= \frac{\{\hat{S}^k, \hat{p}_i\}}{2Vm_e},\\
\hat{J}^{\hat{L}^k}_{i} &= \frac{\{\hat{L}^k, \hat{p}_i\}}{2Vm_e} ,
\end{align}
where $\{\hat{A},\hat{B}\} = \hat{A}\hat{B} + \hat{B}\hat{A}$ is the anti-commutator, $\hat{S}^k$ ($\hat{L}^k$) the $k^{\text{th}}$ component of the spin (orbital) angular momentum operator, $\hat{p}_i$ the $i^{\text{th}}$ component  of the momentum operator, $V$ the volume of the unit cell, and $m_e$ the electron mass.

Using the Kubo linear-response formalism  \cite{Kubo1957,Oppeneer2001,Salemi2021}, $\boldsymbol{\sigma}^{S^k}$ and $\boldsymbol{\sigma}^{L^k}$ can then be computed using
\begin{equation}
\label{eq:LinearResponse}
\begin{split}
\mathcal{A} &= -\frac{ie}{m_e} \int_{\Omega} \frac{d\bm{k}}{\Omega}
\sum_{n\neq m} \frac{f_{n\bm{k}} - f_{m\bm{k}} }{\hbar \omega_{nm\bm{k}}}~
\frac{A_{mn\bm{k}} ~ p_{j,nm\bm{k}} }{-\omega_{nm\bm{k}} + i\tau_{\text{inter}}^{-1}}\\
& ~~~ -\frac{ie}{m_e} \int_{\Omega} \frac{d\bm{k}}{\Omega}
\sum_{n} \frac{\partial f_{n\bm{k}} }{\partial \epsilon}~
\frac{A_{nn\bm{k}} ~ p_{j,nn\bm{k}} }{i\tau_{\text{intra}}^{-1}} \, ,
\end{split}
\end{equation}
where $f_{n\bm{k}}$ is the occupation of Kohn-Sham state $|n\bm{k}\rangle$ with energy $\epsilon_{n\bm{k}}$,  $\Omega$ is the Brillouin zone volume, $p_{j,nm\bm{k}}$ the $j^\text{th}$ component of the momentum-operator ($\hat{\bm{p}}$) matrix element,
and $ \hbar \omega_{nm\bm{k}} = \epsilon_{n\bm{k}} - \epsilon_{m\bm{k}}$. $A^i_{mn\bm{k}}$ stands for a generic matrix element of a generic operator $\hat{A}$; if it is $A_{mn\bm{k}} = \hat{J}^{\hat{S}^k}_{i,mn\bm{k}}$ then $\mathcal{A} = \sigma^{S^k}_{ij}$ while if it is $A_{mn\bm{k}} = \hat{J}^{\hat{L}^k}_{i,mn\bm{k}}$ then $\mathcal{A} = \sigma^{L^k}_{ij}$. The parameters $\tau_{\text{inter}}$ and $\tau_{\text{intra}}$, which are lifetime parameters, are set to $0.4$ eV, a value that is reasonable for metals \cite{Oppeneer2001}. They are effective decay constants modelling interactions with external baths (e.g., phonons). 

It is important to understand that our calculations give the intrinsic parts of the SHE, OHE, SNE, and ONE in the sense that they focus on the Berry curvature related spin and orbital transport coefficients. They are not intrinsic in the sense that the system is considered impurity-free as we use finite lifetime parameters $\tau_{\text{inter}}$ and $\tau_{\text{intra}}$. A further appropriate note at this point is that for the transverse spin and orbital conductivities the intraband part does not contribute. 
Also,  we note that our formulation [Eq.\ (\ref{eq:LinearResponse})] differs from other formulations in which the lifetime broadening appears also in the denominator, i.e., $ (f_{n\bm{k}} - f_{m\bm{k}} )/ [ \omega_{nm\bm{k}}-i\tau_{\rm inter}^{-1}]$, giving thus a denominator of the form $[\epsilon_{n\bm{k}} - \epsilon_{m\bm{k}} + i \delta]^2$, with $\delta= \hbar \tau^{-1}_{\rm inter}$.
The inclusion of the lifetime broadening $\tau_{\text{inter}}$ and $\tau_{\text{intra}}$ in our formulation is done from the beginning of the derivation of the Kubo linear-response formalism as an effective non-Hermitian decay of the first-order correction to the density matrix. Therefore, we argue that the formulation of Eq.\ (\ref{eq:LinearResponse}) is exact in the linear-response framework.

\begin{figure*}[ht!]
  \includegraphics[width=\textwidth]{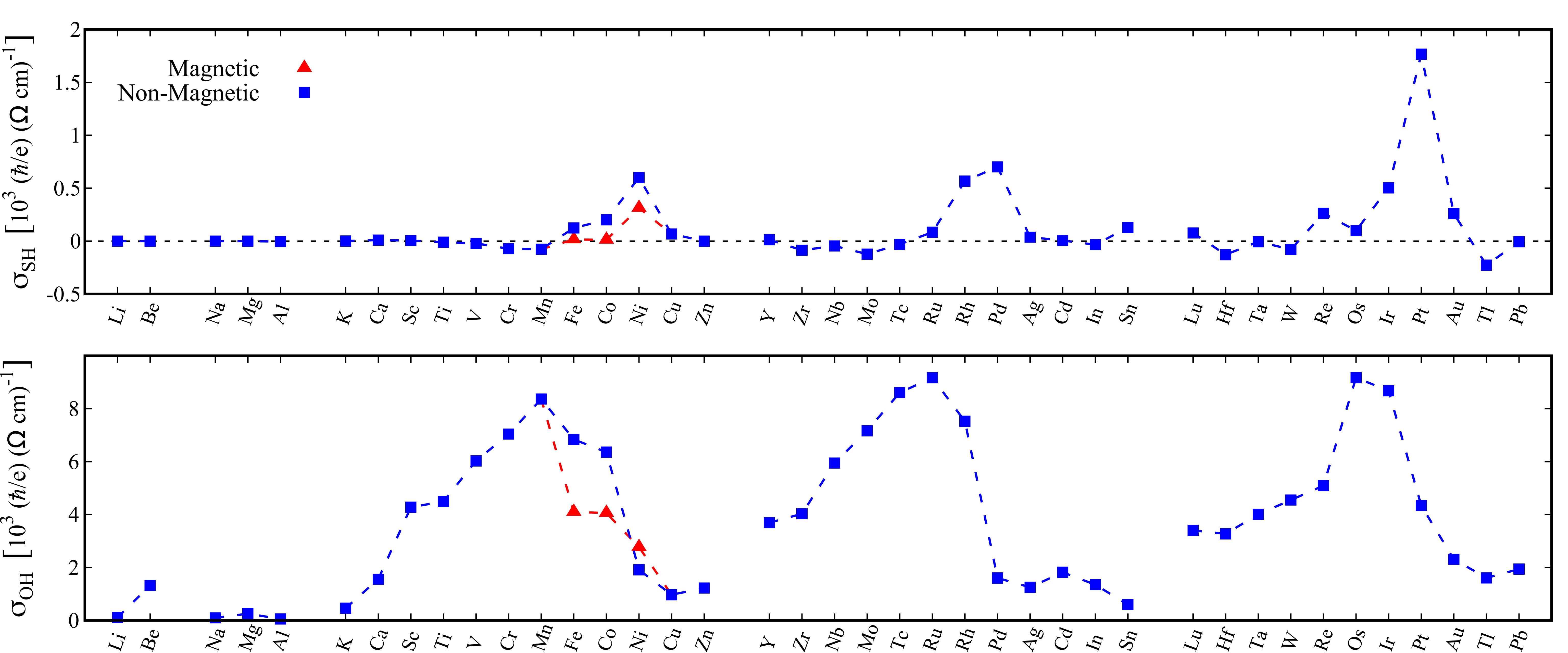}
  \caption{Calculated spin Hall conductivity $\sigma_{\text{SH}}$ (top) and orbital Hall conductivity $\sigma_{\text{OH}}$ (bottom) for a range of nonmagnetic monoatomic metallic materials. For the ferromagnetic 3$d$ elements Fe, Co, and Ni, the values for their magnetic phase are shown as red triangles.}
  \label{fig:SHE_OHE_All_Elements}
\end{figure*}

The conductivities $\SHC$ and $\OHC$ are computed as a function of the electrochemical potential $E$ where $E=0$ corresponds to the Fermi level. This is facilitated by the occupation numbers $f_{n\bm{k}}$ in Eq.~(\ref{eq:LinearResponse}), given by the Fermi-Dirac distribution 
\begin{equation}
f_{n\bm{k}}(E) = \frac{1}{\exp\Big(\frac{\epsilon_{n\bm{k}} - (E_F+E)}{k_B T}\Big) + 1},
\end{equation}
where the electronic potential $E$ is then treated as a variable, allowing us to compute $\SHE(E)$ and $\OHE(E)$. The SNE and ONE transport coefficients $\SNC$ and $\ONC$ can then be computed using the Mott formula [Eq.\ (\ref{eq:Mott_Formula})].

\subsection{Computational Details}
The required equilibrium electronic structures of the monoatomic systems are computed  using the full-potential, all-electron WIEN2k method \cite{Blaha2018}. In this method, the product between the smallest muffin-tin radius $R_{MT}$ and the largest reciprocal vector $K_{\text{max}}$, $RK_{\text{max}} = R_{MT} \times K_{\text{max}}$ is set to appropriate values for each system (see Table~\ref{tab:Structure_List} in the Appendix \ref{Appendix} for more details). The $k$-meshes contain at least $2\, 10^4$ $k$-points. Second, a single-shot calculation using
a refined $k$-mesh of at least $2 \, 10^5$ $k$-points is performed and $\SHC(E)$ and $\OHC(E)$ are computed according to Eq.~(\ref{eq:LinearResponse}). In a post processing step, Eq.~(\ref{eq:Mott_Formula}) is used to compute $\SNC(E)$ and $\ONC(E)$. Structural information on the systems computed in this work can be found in Table~\ref{tab:Structure_List}.

\section{Results}
\label{Results}

\subsection{SHE and OHE}

We first focus our discussion on the SHE and OHE. Figure \ref{fig:SHE_OHE_All_Elements} shows the $\SHE$ (top panel) and $\ONE$ (bottom panel) values, calculated at the Fermi level for all the materials considered. The blue squares show $\SHE$ and $\OHE$ for the nonmagnetic phase while the red triangle shows the results for  ferromagnetic Fe, Co, and Ni. For both $\SHE$ and $\OHE$, we observe a clear trend, where each transition metal row of the Periodic Table (3$d$, 4$d$, and 5$d$) shows a similar pattern.

\begin{figure*}[ht!]
  \includegraphics[width=0.98\textwidth]{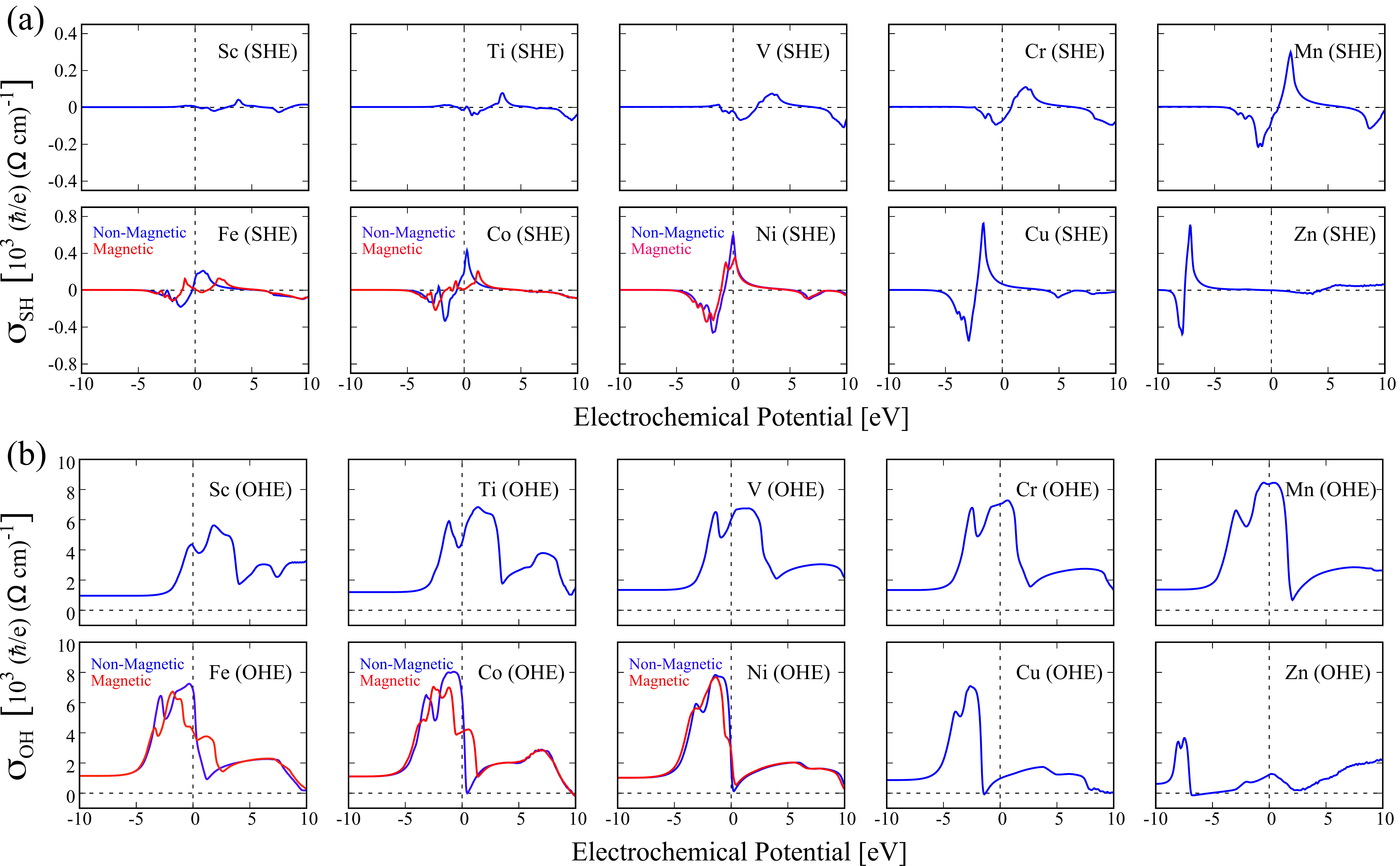}
  \caption{(a) Spin Hall conductivity $\sigma_{\text{SH}}$ and (b) orbital Hall conductivity $\sigma_{\text{OH}}$ computed as a function of the electrochemical potential $E$ for the 3$d$ series. The equilibrium Fermi level is at $E = 0$. Note that the ordinate scales in (a) are different for the two rows, reflecting that $\sigma_{\text{SH}}$ increases with the atomic number $Z$. For Fe, Co, and Ni, results are shown for both the ferromagnetic (red) and nonmagnetic (blue) phases.}
      \label{fig:SHE_OHE_3d}
\end{figure*}

For the SHE, a local maximum is reached for Ni, Pd, and Pt, respectively, for the 3$d$, 4$d$, and 5$d$ series. Those three elements are located in the same column, the group 10 elements of the Periodic Table. These metals have similar electron configurations with 10 electrons in their $s$ and $d$ shells. 
Their $d$ shells are filled to around half of the middle of the second half of the $d$ series (about $d^8$).
Going from the 3$d$ to the $4d$ and to the $5d$ series, one notices that the $\SHE$ increases, which was to be expected since the SHE is known to be a SOC dependent property \cite{Dyakonov1971,Sinova2015} (and the
 SOC strength increases with an increasing atomic number $Z$). 
 The largest SHC value of 1800 $(\hbar/e)\,(\Omega\,\textrm{cm})^{-1}$ is found for Pt, consistent with previous theoretical work \cite{Guo2008,Stamm2017}. The SHC of W is negative, but small, as the calculation here is done for the bcc ($\alpha$) phase and not for the metastable $\beta$ phase, which has a higher SHC \cite{Qian2020}.
 
 It can furthermore be seen that the SHC is very small in the light $sp$ metals (Li to Ca). Elements with a filled $d$ shell, such as Zn and Cd, also have a very small SHC. The SHC of $p$-band metals as In, Sn, and Pb is quite small, too, and only Tl with a large SOC has a somewhat larger SHC (-210 $(\hbar/e)\,(\Omega\,\textrm{cm})^{-1}$).
 Together, this underlines that, to obtain a large SHC, one needs strong SOC as well as $d$-band electrons at the Fermi energy.

For the OHE, a local maximum is reached around the middle of the three $d$ series, specifically, for Mn, Ru, and Os. Mn is located in the middle of the 3$d$ series (group 7 of the Periodic Table, with five $d$ electrons), but Ru and Os are both located in group 8. There is thus a difference for the OHE between the $3d$ series and the $4d$ and $5d$ series, which could be due to the increased spin-orbit coupling when the atomic number $Z$ increases. This gives a larger splitting of $d_{3/2}$ and $d_{7/2}$ subshells and therefore a somewhat different $d$ band filling. Also, while for the SHE maxima the metals Ni, Pd, and Pt have the same crystal structure (face-centered cubic), Mn, Ru, and Os do not have the same crystal structure. Mn is body-centered cubic while Ru and Os are hexagonal close-packed. Importantly, as opposed to the SHE, the OHE does not seem to scale as we proceed from one $d$ series to the other, suggesting that $\OHE$ barely depends on the SOC strength. We will address this later on.

It is instructive to compare our calculations with previous work. Jo \textit{et al.} \cite{Jo2018} computed the OHC of several $3d$ elements. They obtained a maximum OHC of about $9\,10^{3}$ $(\hbar/e)(\Omega\,$cm)$^{-1}$ for Mn; our results are in good agreement with their calculations. Tanaka \textit{et al.} \cite{Tanaka2008} computed the OHC of 4$d$ and 5$d$ elements. They obtained however a  different trend across the series, with a maximum OHC for the 4$d$ series at Mo and for the 5$d$ series at Ir. Also their values for the OHC are in general smaller (about $4\,10^{3}$ $(\hbar/e)(\Omega$\,cm)$^{-1}$ and less). These differences pinpoint  the need for \textit{ab initio} calculations to achieve precise SHC and OHC values.   

Most of the investigated elements are nonmagnetic,  with the exception of Fe, Co, and Ni that are well-known ferromagnets. We see that ferromagnetism (red data points in Fig.\ \ref{fig:SHE_OHE_All_Elements}) has a tendency of reducing the magnitude of both the SHE and OHE, the exception being the OHE for Ni that is slightly higher in the ferromagnetic phase than in the nonmagnetic phase. As discussed below, this effective reduction is due to the spin spitting of the $d$ bands around the Fermi energy, caused by the Stoner instability.

\begin{figure*}[ph!]
  \includegraphics[width=0.98\textwidth]{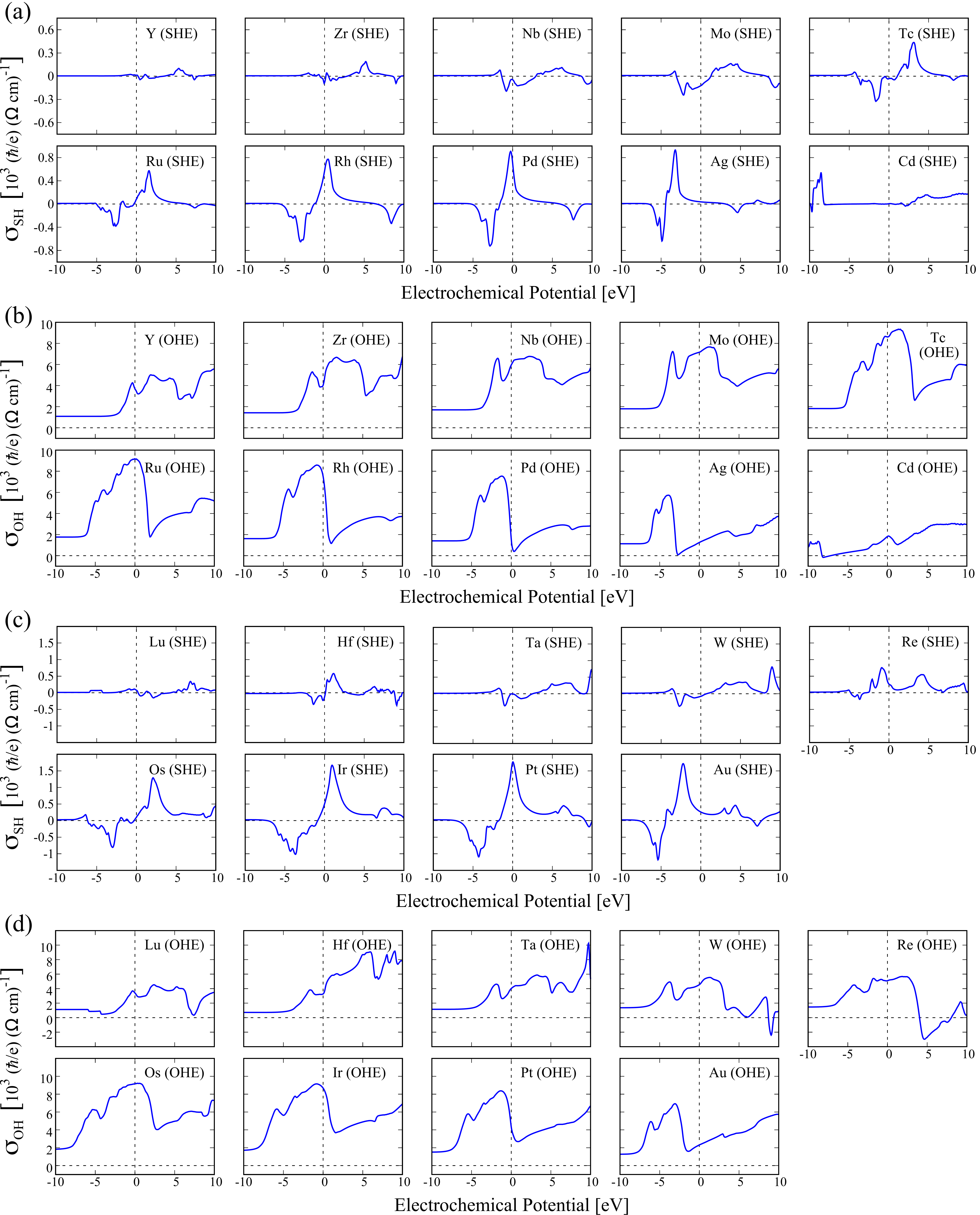}
  \caption{Electrochemical potential dependence of (a) the spin Hall conductivity $\SHE$ for the 4$d$ series, (b) the orbital Hall conductivity $\OHE$ for the 4$d$ series, (c) the spin Hall conductivity $\SHE$ for the 5$d$ series, and (d) the orbital Hall conductivity $\OHE$ for the 5$d$ series. For the 5$d$ series, Hg, which is liquid at room temperature, has not been considered.}
    \label{fig:SHE_OHE_4d_5d}
\end{figure*}

We now compute $\SHE(E)$ and $\OHE(E)$ where $E$ denotes the electrochemical potential (ECP), and $E=0$ is the equilibrium Fermi energy (see Fig.\ \ref{fig:SHE_OHE_3d}). We consider $E$ within the range $E\in[-10; 10]$ eV. For both SHE and OHE, we observe that the shape of $\SHE(E)$ and $\OHE(E)$ is essentially the same across the series, but the central part of the spectrum shifts toward lower energies as the atomic number $Z$ is increased. This clearly shows that the SHE and OHE depend on the amount of $d$-shell filling. For the $\SHE(E)$, the amplitude of the maximum increases as the atomic number $Z$ increases but such behavior is not observed for $\OHE(E)$ for which the maximum seems to be higher when it is close to $E=0$ and smaller when it occurs at higher or lower energies.

For the spontaneous ferromagnets Fe, Co, and Ni the $\SHE$ and $\OHE$ have been computed both for the nonmagnetic and magnetic phases, shown by blue and red curves, respectively. The emergence of ferromagnetism in those metals is well understood: a high density of states around the Fermi energy in the nonmagnetic phase gives rise to a Stoner instability, which in turn leads to an exchange splitting of the $d$ bands in spin-up and spin-down states. This influence of this splitting of the $d$ bands can be observed directly for both the $\SHE(E)$ [Fig.\ \ref{fig:SHE_OHE_3d}(a)] and the $\OHE(E)$ [Fig.\ \ref{fig:SHE_OHE_3d}(b)], where the nonmagnetic curves become split in the spectral range around the Fermi level. This suggests that inherent magnetism arising from a high-density instability at the Fermi energy tends to play against large values of the SHE and OHE.

In a similar fashion, we compute $\SHE(E)$ and $\OHE(E)$ for the 4$d$ and 5$d$ series. As shown in Fig.\ \ref{fig:SHE_OHE_4d_5d}, the same kind of observations as for the 3$d$ series can straightforwardly be made: the shape of both the SHE and OHE is similar for all elements while the position of the maximum tends to be shifted toward lower energies as $Z$ increases. The SHE increases in magnitude as $Z$ increases and the OHE tends to be stronger when the maximum of the spectrum is near the Fermi energy. Particular attention can be paid to Pt, often considered as the best metallic candidate for transverse spin-current generation \cite{MihaiMiron2011,Liu2012review,Hoffmann2013}. The fact that the SHE is largest in Pt stems from the location of the maximum of $\SHE(E)$, which occurs at the Fermi level. Therefore, electronic-structure engineering, such as doping, is not required to achieve a maximal $\SHE(E)$ for Pt. It deserves nonetheless to be noted that Ir and Au have quite similar $\SHE$ maxima, which suggests that doping these metals could bring  their ECP to the maximal SHC position and hence they should yield similar performances as Pt.

\subsection{SOC scaling of SHE and OHE}
Our results, consistent with the literature \cite{Hirsch1999,Guo2008,Sinova2015}, suggest that the strength of SOC plays an important role, especially in the emergence of the SHE. To obtain a better understanding of the interplay of SOC and SHE/OHE, we introduce in our calculation a parameter $\alpha$ ($\alpha \in \Re^+$), the SOC scaling parameter, which is artificially inserted in the Kohn-Sham Hamiltonian $\hat{H}$ such that
\begin{equation}
\hat{H} = \hat{H}_{\text{scRel}} + \alpha \, \hat{H}_{\text{SOC}},
\end{equation}
where $\hat{H}_{\text{scRel}}$ is the scalar-relativistic part of the Hamiltonian and $\hat{H}_{\text{SOC}}$ is the SOC part of $\hat{H}$, i.e.\ $\hat{H}_{\text{SOC}} \propto \hat{\bm{L}} \cdot \hat{\bm{S}}$.
The SOC scaling parameter $\alpha$ controls how strong the SOC is; $\alpha=1$ corresponds to the real SOC in the material. Our calculations are done fully self-consistently, that is, the electronic density $n(\bm{r})$ is computed self-consistently for each $\alpha$ value considered.

\begin{figure}[ht!]
  \includegraphics[width=\linewidth]{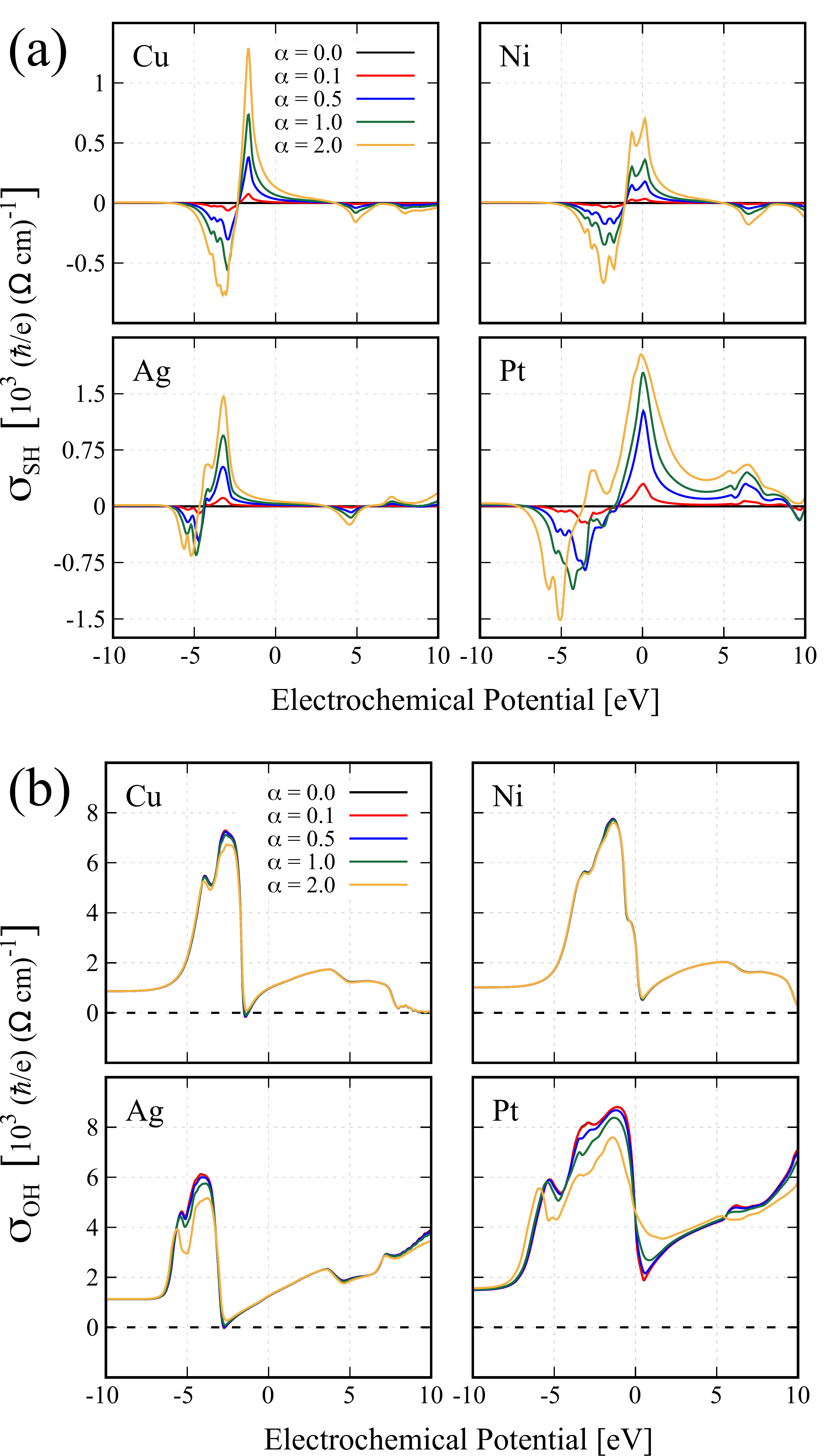}
  \caption{(a) Spin Hall conductivity $\sigma_{\text{SH}}$ and (b) orbital Hall conductivity $\sigma_{\text{OH}}$ computed for different SOC scaling $\alpha$ for Cu, ferromagnetic Ni, Ag, and Pt. The spin Hall conductivity vanishing when we suppress the SOC ($\alpha=0$), while the orbital Hall conductivity remains finite. Increasing $\alpha$ increases $\sigma_{\text{SH}}$ and somewhat decreases $\sigma_{\text{OH}}$ for Ag and Pt. While the scaling appears linear for lighter elements at low $\alpha$, non-linear behavior can clearly be observed for Pt.}
  \label{fig:Pt_Ag_Cu_Ni_SOC_Scaling_Spectrum}
\end{figure}

\begin{figure}[ht!]
  \includegraphics[width=\linewidth]{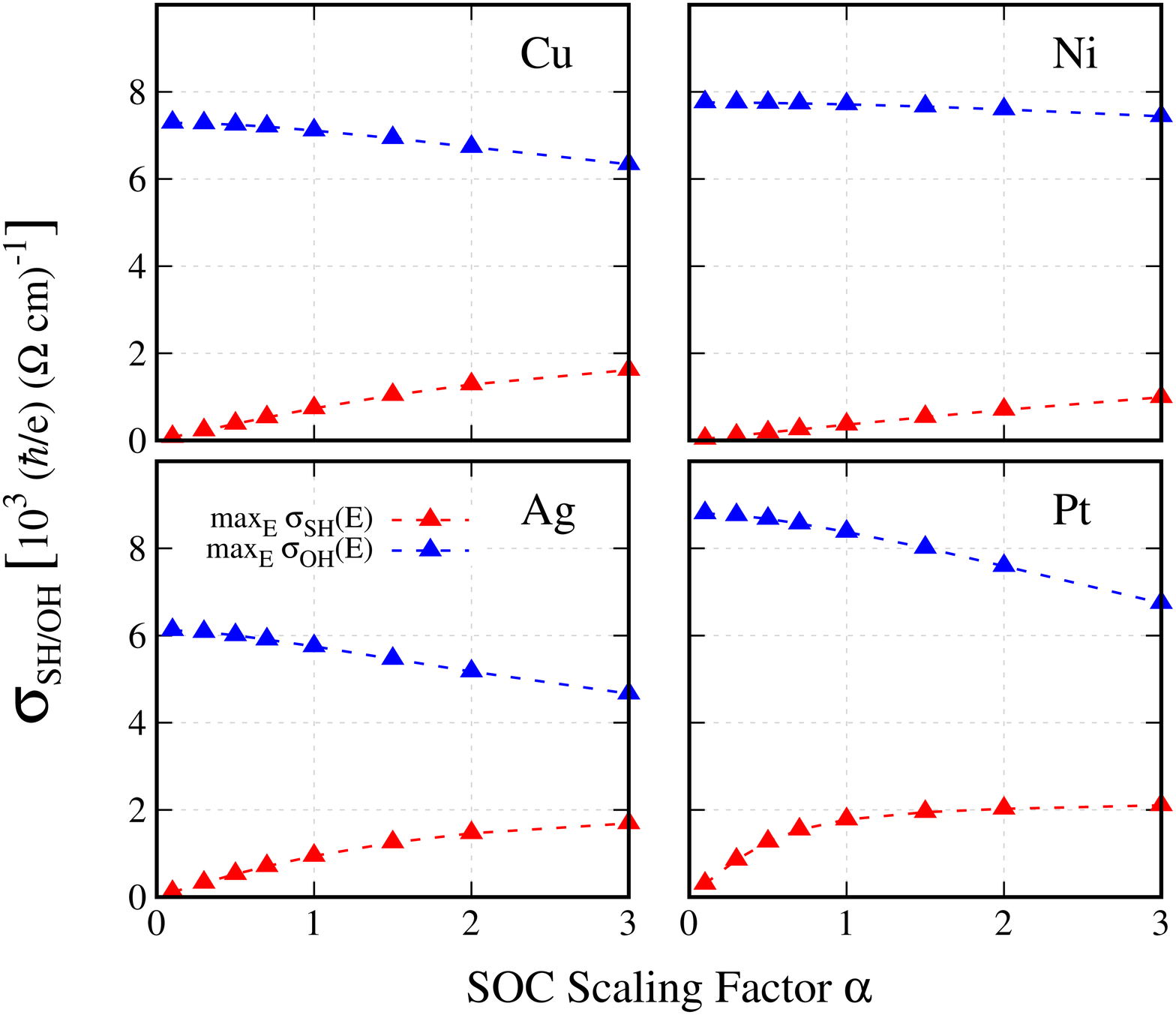}
  \caption{Scaling of the maximum of $\SHE$ (red triangles) and the maximum of $\OHE$ (blue triangles) as a function of the SOC scaling factor $\alpha$ at the peak of the effect ($\max_E \SHE(E)$ and $\max_E \OHE(E)$).}
  \label{fig:SHE_OHE_SOC_Scaling_Max_Alpha}
\end{figure}

\begin{figure}[ht!]
  \includegraphics[width=0.8\linewidth]{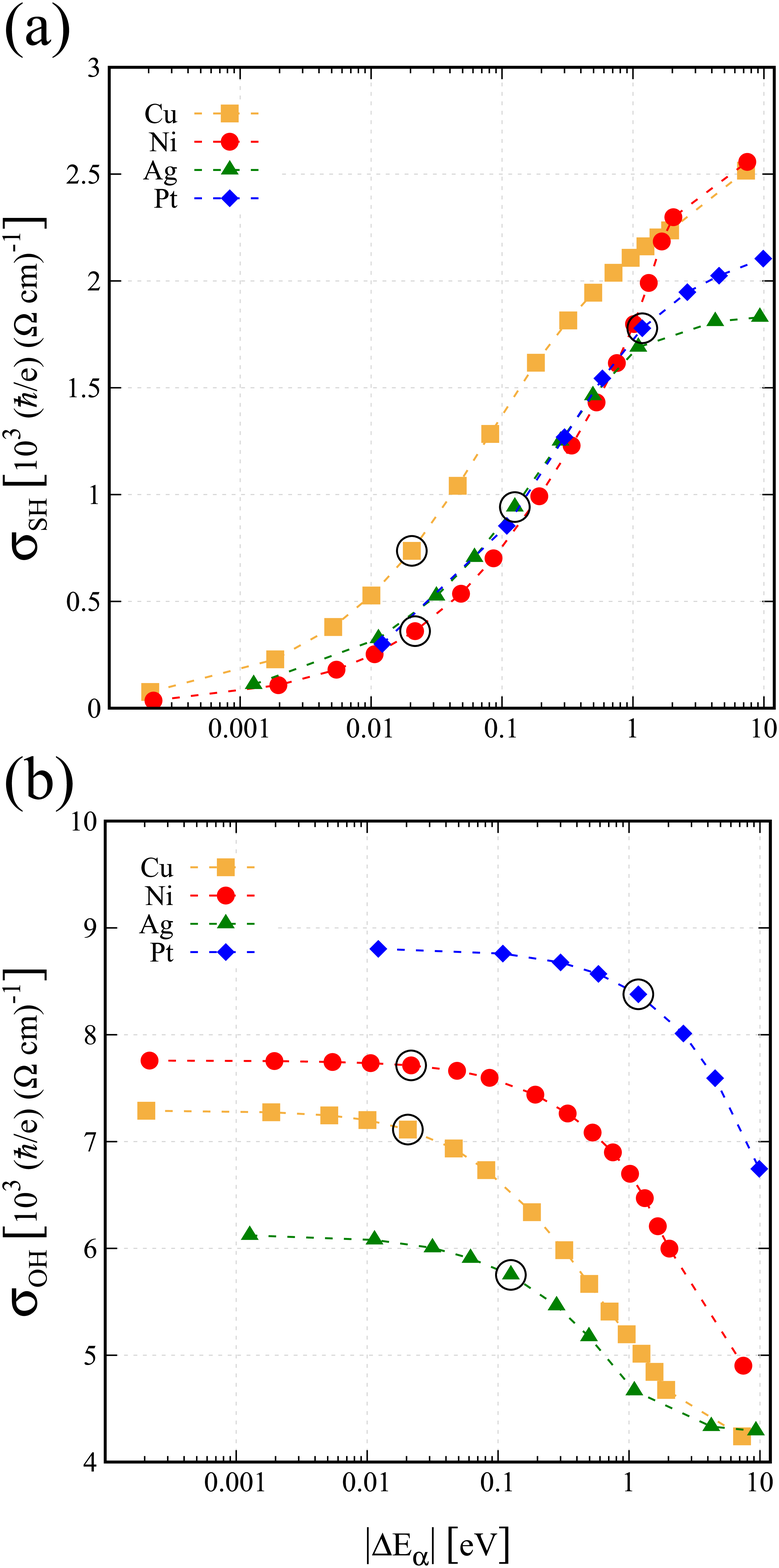}
  \caption{Scaling of (a) the maximum of $\SHE$ and (b) the maximum of $\OHE$ as a function of $|\Delta E_\alpha|$ for Cu (yellow squares), Ni (red circles), Ag (green triangles) and Pt (blue diamonds). The points corresponding to $\alpha = 1$ (``true SOC strength") are circled in black.  For the SHE, the scaling of the effect as a function of $|\Delta E_\alpha|$ follows a similar trend, qualitatively but also quantitatively, for all the elements considered, suggesting that the magnitude of $\SHE$ is mainly determined by the SOC strength rather than by the band structure. For the OHE the considered elements follow qualitatively a similar trend.}
  \label{fig:Pt_Ag_Cu_Ni_SOC_Scaling_Energy}
\end{figure}

We consider four materials, Ni, Cu, Ag, and Pt, such that each row of the $d$ series is considered, taking care of including a magnetic (Ni) and a nonmagnetic (Cu) material for the $3d$ series. The SHE and OHE spectra, $\SHE(E)$ and $\OHE(E)$, for varying $\alpha$ are shown in Fig.\ \ref{fig:Pt_Ag_Cu_Ni_SOC_Scaling_Spectrum}. For all considered materials, no SHE can be observed in the case $\alpha=0$, which is of course to be expected: the lack of coupling between the real-space and the spin-space in $\hat{H}_{\text{scRel}}$ forbids any coupling between the electron momentum (which couples to the external electric field) and its spin angular momentum. For the OHE, in contrast, a finite effect is observed in all cases even when $\alpha=0$.

The $\alpha$ dependencies of the SHE and the OHE are different: while the SHE increases as $\alpha$ increases, the OHE has a weak tendency to decrease in amplitude. The decreasing trend for the OHE is not negligible for Ag and Pt. This observation is different from previous works where it was suggested that the OHE virtually did not depend on SOC strength, or even increased with the SOC strength \cite{Tanaka2008}. We suspect the reason for this difference to be two-fold. First, it is common to include the SOC in a one-shot non-selfconsistent manner, which may shadow any influence of the SOC onto the orbital character of electronic states in the proximity of the Fermi energy. Second, if we were to focus at the variation around $E=0$, we would indeed observe a negligible influence of the SOC strength for all material considered here. The largest variations, both for $\SHE(E)$ and $\OHE(E)$, are observed close to their maximum.

We compute $\max_E \SHE(E)$ and $\max_E \OHE(E)$ as a function of $\alpha$. As shown in Fig.\ \ref{fig:SHE_OHE_SOC_Scaling_Max_Alpha}, one can see that for the considered materials, the OHE follows an opposite trend to the SHE. The trend appears to be linear at low $\alpha$ while larger $\alpha$ leads to saturation and nonlinear scaling. The critical $\alpha$ for which the linear behavior starts to become nonlinear is different for the considered materials. Whereas for Cu and Ni the change of $\SHE$ and $\OHE$ mostly follows a linear trend, non-linear behavior is seen for $\alpha > 1$ for Ag and $\alpha > 0.5$ for Pt. Interestingly, the saturation value for $\SHE$ seems to lie around $\SHE \sim 2000 \, (\frac{\hbar}{e}) (\Omega \, \textrm{cm})^{-1}$, independent of the material considered.

Since the intrinsic SOC tends to scale as $Z^2$ to $Z^4$  \cite{Shanavas2014} we consider the quantity $\Delta E_\alpha$, defined as
\begin{equation}
\Delta E_\alpha = \langle \hat{H}_{\text{scRel}} + \alpha  \hat{H}_{\text{SOC}} \rangle_{n(\bm{r},\alpha)} - \langle \hat{H}_{\text{scRel}} \rangle_{n(\bm{r},0)}
\end{equation}
where $\langle . \rangle_{n(\bm{r},\alpha)}$ refers to the expectation value with respect to the self-consistent density $n(\bm{r},\alpha)$. Note that this definition of $\Delta E_\alpha$ is not equivalent to $\langle \alpha  \hat{H}_{\text{SOC}} \rangle_{n(\bm{r},\alpha)}$ because we take the self-consistent effect of the SOC on the electronic structure into account.

\begin{figure*}
  \includegraphics[width=\textwidth]{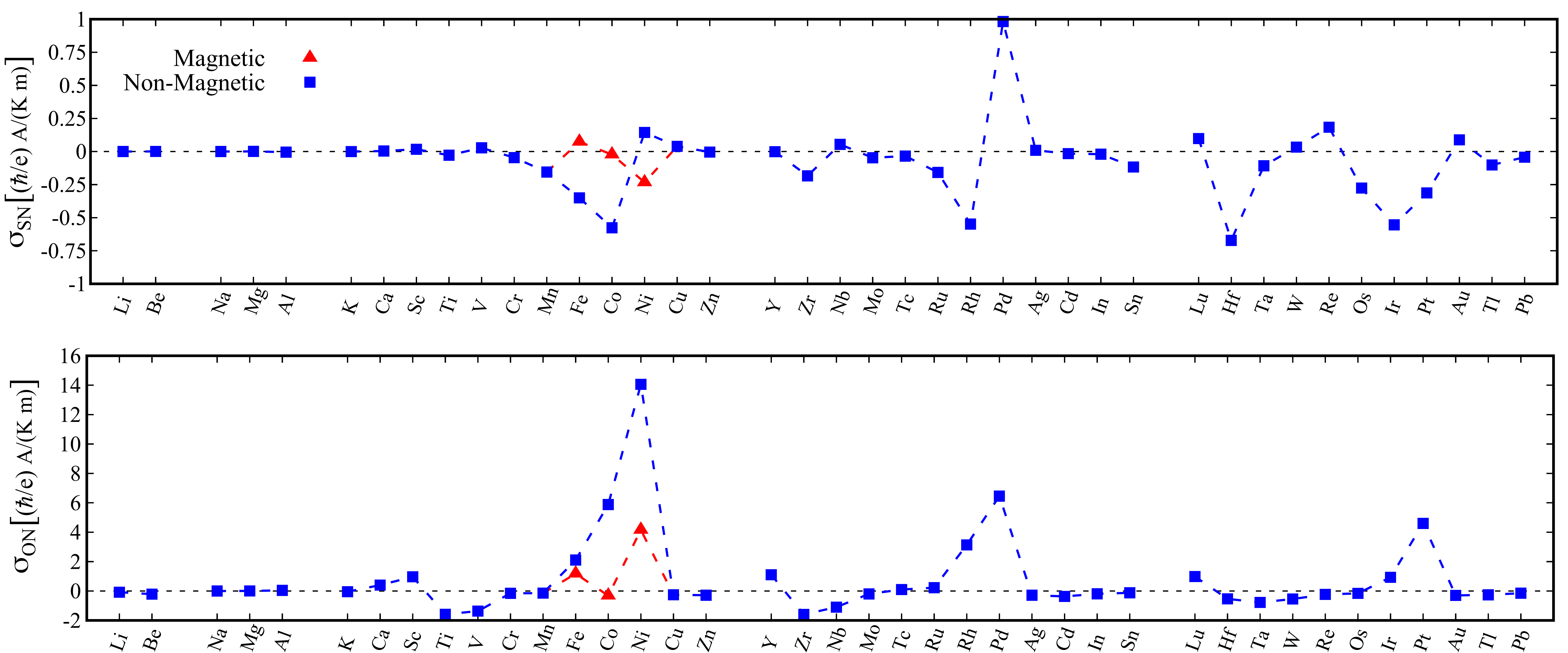}
  \caption{Spin Nernst conductivity $\sigma_{\text{SN}}$ (top) and orbital Nernst conductivity $\sigma_{\text{ON}}$ (bottom) computed for a temperature of $300$\,K for a range of materials. For the ferromagnetic 3$d$ elements Fe, Co, and Ni, results for the ferromagnetic phase are shown as red triangles.}
    \label{fig:SNE_ONE_All_Elements}
\end{figure*}

The scaling of $\max_E \SHE(E)$ and $\max_E \OHE(E)$ with respect to $\Delta E_\alpha$ is shown in Fig.\ \ref{fig:Pt_Ag_Cu_Ni_SOC_Scaling_Energy}. For Cu and Ni, Ag, and Pt,  $\Delta E_\alpha$ is of the order of $10^{-2}$ eV, $10^{-1}$ eV, and $10^0$ eV, respectively, i.e., increasing by an order of magnitude when going from one of the $d$-series to the next one, consistent with the increase of the SOC with $Z$.

 For the SHE [Fig.\ \ref{fig:Pt_Ag_Cu_Ni_SOC_Scaling_Energy}(a)], the $\SHE$ tends to increase rapidly up to $~1$ eV and then starts to saturate. A remarkable point here is that the magnitude of $\SHE$ at a given $\Delta E_\alpha$ is similar for all materials considered, with a slightly higher value for Cu. For Ni, something peculiar happens: for $\Delta E_\alpha < 1$ eV, the behavior of $\SHE$ is similar to Pt and Ag, while for $\Delta E_\alpha > 2$ eV it follows closely the one of Cu. This change of behavior for Ni can be correlated to the decrease of the magnetic moment as a function of $\Delta E_\alpha$. We have observed that for $\Delta E_\alpha \gtrsim 2$ eV, magnetism in Ni disappears.

For $\max_E \OHE(E)$ [Fig.\ \ref{fig:Pt_Ag_Cu_Ni_SOC_Scaling_Energy}(b)], the behavior is opposite to the one of $\max_E \SHE(E)$. For $\Delta E_\alpha < 0.1$ eV, the $\OHE$ is virtually independent to the SOC strength. Beyond this threshold, $\max_E \OHE(E)$ decreases monotonically. Here, Ni's $\max_E \OHE(E)$ does not join the curve of Cu for large $\Delta E_\alpha$ suggesting that presence or absence of magnetism has little influence on the $\OHE$.

\subsection{SNE and ONE}
From the calculated $\SHE(E)$ and $\OHE(E)$ we can compute the thermal $\SNE$ and $\ONE$ using the Mott formula, Eq.\ (\ref{eq:Mott_Formula}). Note that since the Mott formula 
is valid only around the Fermi level, one cannot extract the ECP profile of the SNE and ONE. We choose $T=300$\,K and evaluate the SNE and ONE for this temperature. In Fig.~\ref{fig:SNE_ONE_All_Elements}, we display the computed $\SNE$ and $\ONE$ for all the elements considered in this work. Similarly to the $\SHE$ and $\OHE$, we find that the orbital Nernst values are systematically much larger than their spin counterparts. The effect of ferromagnetism on the $\SNE$ and $\ONE$ is also quite similar to their Hall counterparts, with a substantial increase of the $\SNE$ and $\ONE$ when magnetism is artificially turned off for Fe, Co, and Ni. A further similarity with the Hall conductivities is that the Nernst conductivities are very small for $sp$ metals.

Differences with respect to the Hall conductivities are also noticeable. First, as opposed to the $\SHE$ and $\OHE$, the maximum value of the $\SNE$ is obtained for Pd, a $4d$ element. This shows that although the maximum values of the $\SHE$ relate directly to the increase of the SOC, there is no clear scaling of the $\SNE$ as it relates to the derivative of $\SHE$. 
Conversely, the ONE does reveal a trend across the $d$ series. The maximum of the (nonmagnetic) ONE occurs for the group 10 elements, Ni, Pd, and Pt. Also, there is a change in sign that occurs around the middle of each $d$ series, with the early $d$ elements having negative ONEs. Like the OHE, the ONE does not depend on the SOC, but in addition, the narrowness of the $d$ bands plays a role, and Ni has the most correlated  3$d$ bands of the group 10 elements.  Although the orbital Nernst effect has not yet been observed, our results suggest that giant ONEs could be present in correlated materials having narrow bands with a strong variation of the $d$ or $f$ states density close to the Fermi energy. 

Experimental and theoretical investigations of the spin Nernst effect were reported for several materials \cite{Tauber2012,Wimmer2013,Geranton2015,Meyer2017,Sheng2017,Bose2018}, see Ref.\ \cite{Bose2019} for a recent review.
Our results are in overall agreement with those of G\'eranton \textit{et al.}\ \cite{Geranton2015}, who gave computed values of the intrinsic Nernst coefficient for Rh, Pd, Ir and Pt, but our values are roughly a factor of two smaller. Our value for Pt is however in good agreement with the calculated value of Meyer \textit{et al.}\ \cite{Meyer2017}. Differences between \textit{ab initio} calculated values could be due to the choice of lifetime broadening. A smaller lifetime broadening will lead to sharper features in the SHC and OHC spectra and thus modify the SNC and ONC.

\section{Conclusions}
In this work, we investigated the intrinsic $\SHE$ and $\OHE$ as a function of the electrochemical potential $E$ for 40 monoatomic elements. We showed that for the $d$-elements, the qualitative shapes of the $\SHE(E)$ and $\OHE(E)$ spectra are  similar for the elements in a specific $d$ series, with the relative position of $E=0$ for each element strongly dependent on the filling of the $d$-shell. 
For the $\OHE(E)$ spectra, maximum values are obtained when the $d$-band is roughly half filled. The $\SHE$ becomes maximal when the Fermi level falls in the middle of the second half of the $d$ series, which happens for (nonmagnetic) Ni, Pd, and Pt.
We also found that magnetism (in Fe, Co, and Ni) tends to reduce $\SHE(E)$ and $\OHE(E)$, by splitting the $d$-states away from the Fermi level.

We furthermore considered the influence of SOC on $\SHE(E)$ and $\OHE(E)$. The $\OHE(E)$ is obtained even when SOC is turned-off, consistent with previous reports \cite{Tanaka2008,Go2018}. We also showed that while it is true that $\OHE(E)$ depends less on the SOC than the $\SHE(E)$, nonetheless, a non-negligible influence is observed. Analyzing the SOC influence using $\Delta E_\alpha$ allowed us to compared elements from the $3d$, $4d$, and $5d$ series on an equivalent footing. Artificially increasing the SOC increases the $\SHE$, but this scaling  saturates as $\Delta E_\alpha$ increases, suggesting some kind of limit to the intrinsic Hall effect.

For the $\SNE$ and $\ONE$, we presented a survey of their theoretical values for the here considered materials. As we showed, the orbital Nernst effect is about one order of magnitude larger than the spin Nernst effect. This is notably for the lifetime broadening adopted here ($\hbar \tau^{-1}_{\rm inter} = 0.4$ eV) and even larger differences can be expected for smaller lifetime broadening.

Our work emphasizes that the orbital contributions, both for the Hall and Nernst effect, should be quite important. Encouraging reports of observations of orbital currents, orbital torque, and the OHE have appeared recently \cite{Ding2020,Lee2021a,Lee2021b}.
The orbital Nernst effect in metals has so far not been detected. A magnonic equivalent of the orbital Nernst effect was proposed recently to exist for magnetic insulators \cite{Zhang2020b}.
There are thus still questions about the nature of orbital transport and its direct experimental observation that remain. It is however interesting to see that the orbital part, which has previously been discarded, is not only far from being negligible but actually appears dominant. Since the orbital Hall and Nernst effects are present without SOC, but not the SHE and SNE, the latter quantities arise from the former ones through the spin-orbit interaction. Large orbital effects could thus be harvested for lighter 3$d$ and 4$d$ metals and compounds \cite{Tanaka2008,Jo2018}, in place of the heavy metals Pt, Ta, and W that are favorable for large SHE.  Hence, we anticipate that our work will contribute and stimulate research in the emergent field of orbitronics.

\begin{acknowledgments}

We thank Marco Berritta for valuable discussions. This work has been supported by the Swedish Research Council (VR) and the Swedish National Infrastructure for Computing (SNIC) (Grant No.\ 018-05973). This work has been funded by the European Union’s Horizon2020 Research and Innovation Programme under FET-OPEN Grant agreement No.\ 863155 (s-Nebula). The calculations were performed at the PDC Center for High Performance Computing and the Uppsala Multidisciplinary Center for Advanced Computational Science (UPPMAX).
\end{acknowledgments}

\appendix
\section{Table of lattice parameters}
\label{Appendix}
\begin{table*}[hp!]
\begin{ruledtabular}
\vspace{2cm}
\centering
\caption[justified]{List of elements considered in this work with their lattice structure and lattice parameters. The structures considered are either face-centered cubic (fcc, $a=b=c$ and $\alpha=\beta=\gamma=\frac{\pi}{2}$), body-centered cubic (bcc, $a=b=c$ and $\alpha=\beta=\gamma=\frac{\pi}{2}$) or hexagonal close-packed (hcp, $a=b\neq c$ and $\alpha=\beta=\frac{\pi}{2}, \gamma = \frac{2\pi}{3}$). The column $RK_\text{max}$ refers to the product between the smallest muffin-tin radius $R_{MT}$ and the largest reciprocal vector $K_{\text{max}}$, and is an important parameter for WIEN2k \cite{Blaha2018} calculations.\\
}
\label{tab:Structure_List}
\begin{tabular}{cccccccccccc}
Element & ~~$a$ [\AA] & ~~$c$ [\AA]~~ & ~~ Structure ~~ & $RK_\text{max}$ & \hspace{2cm}& & Element & ~~$a$ [\AA] & ~~$c$ [\AA]~~ & ~~ Structure ~~ & $RK_\text{max}$  \\
\cline{1-5}
\cline{7-12}
\\
Li    &    3.51    &        &    bcc    &  4.5 &   \hspace{1cm}&    &    Mo    &    3.15    &        &    bcc &  7.5\\[0.1cm]

Be    &    2.29    &    3.58    &    hcp    &    5.0 &   \hspace{1cm}&    &    Tc    &    2.73    &    4.39    &    hcp &  8.0\\[0.1cm]

Na    &    4.29    &        &    bcc    &    6.5 &   \hspace{1cm}&    &    Ru    &    2.71    &    4.28    &    hcp&  8.0\\[0.1cm]

Mg    &    3.21    &    5.21    &    hcp    &    6.5 &   \hspace{1cm}&    &    Rh    &    3.80    &        &    fcc&  8.0\\[0.1cm]

Al    &    4.05    &        &    fcc    &    6.5 &   \hspace{1cm}&    &    Pd    &    3.89    &        &    fcc&  8.0\\[0.1cm]

K    &    5.33    &        &    bcc    &    6.5 &   \hspace{1cm}&    &    Ag    &    4.09    &        &    fcc&  8.0\\[0.1cm]

Ca    &    5.59    &        &    fcc    &    6.5 &   \hspace{1cm}&    &    Cd    &    2.98    &    5.62    &    hcp&  8.0\\[0.1cm]

Sc    &    3.31    &    5.27    &    hcp    &    7.5 &   \hspace{1cm}&    &    In    &    3.25    &    4.95    &    hcp&  8.0\\[0.1cm]

Ti    &    2.95    &    4.69    &    hcp    &    7.5 &   \hspace{1cm}&    &    Sn    &    5.83    &    3.18    &    hcp&  8.0\\[0.1cm]

V    &    3.03    &        &    bcc    &    7.5 &   \hspace{1cm}&    &    Lu    &    3.50    &    5.55    &    hcp&  8.0\\[0.1cm]

Cr    &    2.91    &        &    bcc    &    7.5 &   \hspace{1cm}&    &    Hf    &    3.20    &    5.05    &    hcp&  8.0\\[0.1cm]

Mn    &    3.51    &        &    fcc    &    8.0 &   \hspace{1cm}&    &    Ta    &    3.30    &        &    bcc&  8.0\\[0.1cm]

Fe    &    2.87    &        &    bcc    &    8.0 &   \hspace{1cm}&    &    W    &    3.17    &        &    bcc&  8.0\\[0.1cm]

Co    &    2.51    &    4.07    &    hcp    &    8.0 &   \hspace{1cm}&    &    Re    &    2.76    &    4.46    &    hcp&  8.0\\[0.1cm]

Ni    &    3.52    &        &    fcc    &    8.0 &   \hspace{1cm}&    &    Os    &    2.73    &    4.32    &    hcp&  8.5\\[0.1cm]

Cu    &    3.61    &        &    fcc    &    8.0 &   \hspace{1cm}&    &    Ir    &    3.84    &        &    fcc&  8.5\\[0.1cm]

Zn    &    2.66    &    4.95    &    hcp    &    8.0 &   \hspace{1cm}&    &    Pt    &    3.92    &        &    fcc&  8.5\\[0.1cm]

Y    &    3.65    &    5.73    &    hcp    &    7.5 &   \hspace{1cm}&    &    Au    &    4.08    &        &    fcc&  8.5\\[0.1cm]

Zr    &    3.23    &    5.15    &    hcp    &    7.5 &   \hspace{1cm}&    &    Tl    &    3.46    &    5.52    &    hcp&  8.5\\[0.1cm]

Nb    &    3.30    &        &    bcc    &    7.5 &   \hspace{1cm}&    &    Pb    &    4.95    &        &    fcc&  8.5\\[0.1cm]
\end{tabular}
\end{ruledtabular}
\end{table*}

\bibliographystyle{apsrev4-2}
\bibliography{Bibliography}

\end{document}